\documentclass[pre,showpacs,twocolumn]{revtex4}
\usepackage{epsfig}

\usepackage{amsmath}

\newcommand{\gl}[1]{Eq. (\ref{#1})}

\def\gtrless{\raise2.5pt\hbox{$>$}\llap{\lower2.5pt\hbox{$<$}}}
\def\gtrapprox{\raise2.5pt\hbox{$>$}\llap{\lower2.5pt\hbox{$\approx$}}}
\def\lessapprox{\raise2.5pt\hbox{$<$}\llap{\lower2.5pt\hbox{$\sim$}}}

\newcommand{\vek}[1]{{\bf #1}}

\begin{document}

\title{ Light-Scattering by Longitudinal phonons in Molecular 
Supercooled Liquids I: 
Phenomenological Approach}
\author{R.M. Pick$^{(1)}$, 
 T.~Franosch$^{(2)}$, 
A.~Latz$^{(3)}$, 
C. Dreyfus$^{(4)}$
}
\affiliation{
$^{(1)}$UFR 925, UPMC, Paris, France;
$^{(2)}$Hahn-Meitner Institut, D-14109 Berlin, Germany;
$^{(3)}$Technische Universit{\"a}t Chemnitz, D-09107 Chemnitz, Germany;
$^{(4)}$Physique des Milieux Condens{\'e}s, UPMC, Paris, France }
\date{\today} 

\begin{abstract}
We derive expressions for the intensity of the Brillouin polarized spectrum
of a molecular liquid formed of axially symmetric molecules. 
These expressions take into account both the molecular dielectric anisotropy
and the modulation of the local polarisability by density fluctuations. 
They also incorporate all the retardation effects
which occur in such liquids. We show that the spectrum
splits into a $q$-independent rotational contribution and 
$q$-dependent term, which reflects the propagation of longitudinal phonons.
In the latter, the two light scattering mechanisms enter on an equal
footing and generate three scattering
channels. We study the influence of the two new channels and show 
that they may substantially modify the Brillouin line-shape 
when the relaxation time of the supercooled liquid and the phonon period 
are of the same order of magnitude.

\end{abstract}

\maketitle

%\twocolumn

\section{Introduction}
The theory of inelastic, low frequency, light scattering in molecular liquids
has gone through different episodes over the years. 
The presence of a narrow inelastic doublet in the light scattering
spectrum due to propagating sound waves was predicted
by Brillouin in 1922 \cite{Brillouin1922}. Its first observation 
 was reported in 1930 by Gross, who 
noticed that the spectrum consisted of the Brillouin doublet
superimposed on a central Rayleigh line \cite{Gross1930}. An 
explanation of this Rayleigh-Brillouin triplet, based on the 
equations of macroscopic hydrodynamics, was suggested by Landau and Placzek 
in 1934 \cite{Landau1934} and was given in complete form by Mountain in 1966
\cite{Mountain1966}. This hydrodynamic description did not take into account 
the nature of the constituents of the liquid. Yet,
it can play an important role and typical molecular effects have been detected
on those low frequency spectra ($\lessapprox 50 $ GHz). Let us briefly
mention three of them.

One takes place with molecules that are non-spherical tops.
In that case, both the molecular polarisability tensor and the tensor
of inertia are anisotropic and they can reasonably be considered to 
depend only on the molecular orientation. The (collective)
orientational dynamics of the molecules can thus be detected through the
corresponding fluctuations of the anisotropic part of the dielectric tensor.
This corresponds to a central peak
with a line shape  independent both of the scattering vector
and of the polarisation of the incident and scattered beams. 
Conversely,
the tensorial character of the detection mechanism (anisotropic
part of the local  dielectric tensor) imposes specific relationships 
between the intensities of the spectra when recorded in different 
geometries. In particular, when, as is usually the case, dipole 
induced dipole effects 
(DID) can be neglected \cite{Murphy1999},
 $I_{VV}(\omega) = \frac{4}{3} I_{VH}(\omega)$.
Here the first index corresponds to a vertical (V) polarisation  
of the incident
beam, and the second index to a vertical (V) or horizontal (H)
polarisation of the scattered beam, the scattering plane being 
horizontal. 

The second effect is the possible coupling of the density fluctuations 
with some internal degree of freedom of the molecule, having  a temperature
dependent lifetime, $\tau(T)$. Mountain \cite{Mountain1966} showed that,
when $\omega_B \tau(T)\gg 1$, where $\omega_B$ is the longitudinal 
phonon frequency 
\footnote{Note that, in the whole paper, we use the term 'frequency' as an 
abbreviation for 'circular frequency', $\omega = 2\pi \nu$.}, 
an additional central peak appears in the VV spectrum, 
which can be formally interpreted as the introduction of a 
retardation effect in the bulk viscosity of the liquid. 

A third effect showed up in the study of the VH spectrum of viscous
molecular liquids formed of anisotropic molecules. A wave-vector dependent 
feature was detected, which altered the line-shape of the central peak at 
moderate viscosities \cite{Starunov1967} and transformed into an 
underdamped transverse phonon spectrum for higher viscosities 
\cite{Bezot1975}. 
It was rapidly recognised 
that the additional spectrum was, indeed, due to a transverse, diffusive 
\cite{Keyes1971}, or propagative \cite{Bezot1975} excitation:
any local shear motion inside
the liquid couples to the mean local molecular orientation, 
and renders the latter  anisotropic.
This corresponds to   a similar change of the mean molecular
polarisability tensor. 
The detection mechanism of these shear modes is
thus, as for the first effect, a change in the molecular orientations. 

To rationalise the results obtained in VH light scattering experiments, 
some of the present authors (C.D. and R.M.P) and co-workers,
 \cite{Drey1,Drey2} 
here after referred to as [I], proposed to somewhat generalise the
usual hydrodynamic equations by incorporating through phenomenological 
arguments, in a systematic way:

a) the coupling of the shear deformation to a local mean orientation
of the molecules;

b) a retardation effect in each term corresponding to the damping 
 of a variable.

Concentrating on the case of axially symmetric molecules, they characterised
each of them by the orientation of its axis $\hat{\vek{u}}$, with
polar angles $\theta, \phi$, represented by $P(\theta,\phi,\vek{r},t)$
the local probability density 
of finding $\hat{\vek{u}}$ in that direction, and 
defined a set of orientational density variables, $Q_{ij}(\vek{r},t)$ 
by:
\begin{equation}
Q_{ij}(\vek{r},t)=  \int \! \sin \theta 
d\theta d\phi P(\theta,\phi,\vek{r},t) 
[ \hat{u}_i \hat{u}_j 
- \frac{1}{3} \delta_{ij} ]  \, ,
\end{equation}
where latin indices $i,j,..$ represent cartesian components. 
The set of $Q_{ij}(\vek{r},t)$ forms a symmetrical,
traceless second rank tensor. 

It was proposed in [I] that the hydrodynamic equations pertinent to the
case of a such a molecular viscous liquid would consist, after linearisation,
of the two conservation laws: 
\begin{subequations}
\begin{equation}\label{masscons}
\dot{\rho}(\vek{r},t) + \partial_k J_k(\vek{r},t) = 0 \, ,
\end{equation}
\begin{equation}\label{currcons}
\dot{J}_k(\vek{r},t) = \partial_l \sigma_{kl}(\vek{r},t)
\, ,
\end{equation}
\end{subequations}
where $\rho(\vek{r},t)$ is the mass density, $\vek{J}(\vek{r},t)$ is the
mass current density. Furthermore, they suggested the 
constitutive
equations for the stress tensor:
\begin{equation}\label{constitutive}
\sigma_{ij} = 
(-\delta P + \eta_b \otimes \partial_k v_k ) \delta_{ij} + \eta_s \otimes 
\tau_{ij}
- \mu \otimes \dot{Q}_{ij} \, ,
\end{equation}
and for the orientation:
\begin{equation}\label{Qosci}
\ddot{Q}_{ij} = - \omega_0^2 Q_{ij}
- \Gamma' \otimes \dot{Q}_{ij} 
+ \Lambda' \mu \otimes \tau_{ij} \, .
\end{equation}
Here the momentum density $\vek{J}$ is related to the velocity field 
$\vek{v}$ via the mean mass density $\rho_m$:
\begin{equation}
J_i(\vek{r},t) = \rho_m  v_i(\vek{r},t) \, .
\end{equation}
The strain rate  $\tau_{ij}(\vek{r},t)$ is a second rank 
symmetric and traceless tensor
defined locally by:
\begin{equation}
\tau_{ij} = \partial_j v_i + \partial_i v_j 
- \frac{2}{3} \delta_{ij} \partial_k v_k \, .
\end{equation}
The pressure change $\delta P(\vek{r},t)$
is 
related to the instantaneous
mass density change $\delta \rho(\vek{r},t)$ by:
\begin{equation}\label{pressure}
\delta P(\vek{r},t) = c^2 \delta \rho(\vek{r},t) \, ,
\end{equation}
where $c$ is the relaxed sound velocity.
The retarded couplings are given in terms  of 
 $\eta_b, \eta_s, \mu$ and $\Gamma'$ and are, respectively, the bulk 
and shear viscosities, the rotation-translation coupling and 
the orientational relaxation functions, the symbol $\otimes$ standing 
for a convolution product in time.
Finally, $\omega_0$ is the libration frequency of the axial molecules
and $\Lambda'$ is the rotation-translation coupling constant, a quantity that
takes into account the fact that $\rho$ and $Q_{ij}$
have different dimensions.

Noting that, for motions that do not involve a local density change, 
the local dielectric tensor can be written, neglecting DID effects, as:
\begin{equation}\label{epsQonly}
\delta \epsilon_{ij}(\vek{r},t) = 
b Q_{ij}(\vek{r},t) \, ,
\end{equation}
it was shown in [I] that the preceeding set of equations leads 
to an expression of the intensity that describes the complete thermal 
evolution of the VH spectrum of a supercooled molecular liquid.

Recently, another of the present authors (A.L.) and co-worker \cite{Latz2001}
 made 
an additional step. Within the framework of the Molecular Mode Coupling 
Theory (MMCT) \cite{Schilling1997},
 they expressed the complete dynamics of a system of 
linear molecules characterised by the position of their center of mass and 
their orientation. They showed that, in the $\vek{q} \to 0$ 
limit, both the correlation functions of the density fluctuations
and of the orientation fluctuations contributed to the light scattering
mechanism in the VV geometry. This result stressed the existence of a 
coupling between these two variables for a longitudinal phonon. 
Nevertheless, 
the expression they obtained for the VV intensity 
did not make clear two aspects. One was the separation of the 
intensity into a $q$-independent term, representing a pure rotational 
dynamics, and a hydrodynamic contribution. The second was that the latter
could be factorized into two parts: the longitudinal phonon propagator
and a phonon-photon scattering term in which both the 
density and the orientational fluctuations enter on an equal footing.
Conversely, in another paper published soon after by two of the 
present authors 
(T.F. and A.L.) and co-workers \cite{Franosch2001}, 
this separation into 
two different contributions and the emergence of 
several scattering  channels was made apparent. 
Yet, the technique used in \cite{Franosch2001} did not imply any 
specific light scattering mechanism; it was thus not possible to predict
from that paper which terms would dominate the spectrum 
for a given physical system. 

The present series of papers aims at completing the picture of 
low frequency scattering in viscous molecular liquids through the
combination of two different aspects. 
On the one hand, we shall present a full length, first principle 
derivation of Eqs. (\ref{constitutive},\ref{Qosci}) 
through a Zwanzig-Mori 
approach and obtain from this technique the corresponding
Onsager conditions which, when
met by the relaxation functions,
 guarantee the Brillouin spectra to have 
a positive value whatever the frequency. On the other hand,
we shall complete the results  obtained in [I] by deriving,
using the same phenomenological equations, expressions for the
intensities which can be measured in the VV and HH scattering geometries;
those are the geometries generally used to study longitudinal 
phonons. The results will be completely expressed with the help
of the different quantities entering
Eqs. (\ref{constitutive},\ref{Qosci}) and a generalisation of 
\gl{epsQonly} which includes density fluctuations.

This second series of results have immediate implications for the
experimentalist and they can be derived through elementary 
algebraic techniques.  The formal proof of the 
validity of the phenomenological equations and of the conditions under
which the spectra are positive requires the use of more elaborate tools.
In order to make the present results accessible to as large an audience 
as possible, we found it useful to reverse the logical order and to split 
our presentation into two consecutive papers. 
The second one (Part II) will give a complete derivation of these equations
and of their consequences, only briefly sketched in \cite{Latz2001},
 as well 
as a comparison between the results obtained 
through 
the phenomenological approach and through the more abstract technique of 
\cite{Franosch2001}. 
The phenomenological part of our work (Part I) is organised 
as follows. 

In Section II, we will derive the expressions for the intensity obtained
in VV and HH experiments. Section III will discuss the changes in the
line-shape of the longitudinal phonons
that can be expected when the coupling of the rotation of the 
molecules to the longitudinal phonons is taken into account in the expression
of the dielectric fluctuation. 
Finally, a brief summary of 
these results, a comparison with the expressions previously obtained for
the
VH intensity [I] and additional remarks will be presented in Section IV.

\section{The Longitudinal Phonon Brillouin Scattering Problem}

In a light scattering experiment, the incident laser may be characterised 
by the  amplitude of the electric field 
$E_i$, and its polarization $\hat{\vek{e}}_i$. The spatial and temporal 
variation of the electric field $\vek{E}_{\rm i}(\vek{r},t) = E_{\rm i}
 \hat{\vek{e}}_{\rm i}
\exp i (\vek{k}_{\rm i} 
\cdot \vek{r} - \omega_{\rm i} t)$ is given in terms of the wave vector 
$\vek{k}_{\rm i}$ and the frequency $\omega_{\rm i}$.
 From the corresponding quantities of the
scattered beam, $\hat{\vek{e}}_{\rm f}, \vek{k}_{\rm f}, \omega_{\rm f}$,   
one obtains insight into  the fluctuations of the sample that 
occur at the scattering  vector $\vek{q} = \vek{k}_{\rm i} 
- \vek{k}_{\rm f}$ at the
frequency shift $\omega = \omega_{\rm i} - \omega_{\rm f}$. 
The thermal fluctuations of the dielectric tensor 
$\delta \epsilon_{ij}(\vek{r},t)$ can be decomposed into its spatial 
Fourier components:
\begin{eqnarray}
\delta \epsilon_{ij}(\vek{q},t) = \int d^3 \vek{r} 
\, \, \delta \epsilon_{ij}(\vek{r},t) \exp(i \vek{q} \cdot \vek{r}) \, ,
\end{eqnarray}
a notation that we will use also for other quantities 
for the rest of this paper. This dielectric 
modulation represents a momentary grating by  which  the  laser light is
scattered. Due to the polarisation of the incident beam 
and of 
the analyser for the scattered one, the detector collects only fluctuations
corresponding to the projection of the dielectric fluctuations onto the
two polarisations:
\begin{equation}\label{epsfi}
\delta \epsilon_{\rm fi}(\vek{q},t) = \hat{e}_{{\rm f} k} 
\delta \epsilon_{kl}(\vek{q},t) 
\hat{e}_{{\rm i}k} \, .
\end{equation}

Since the fluctuations readily disappear, there is a corresponding
frequency shift, $\omega$, leading to a total scattered intensity:
\begin{equation}\label{Ifi}
I_{\rm fi}(\vek{q},\omega) = \int_0^\infty dt 
\langle \delta \epsilon_{\rm fi}(\vek{q},t) 
\delta \epsilon_{\rm fi}^0(\vek{q})^{*} 
\rangle \cos ( \omega t) \, ,
\end{equation}
with the notation $\delta \epsilon^0_{\rm fi}(\vek{q}) = 
\delta \epsilon_{\rm fi}(\vek{q},t=0)$  
and similarly for  other quantities. Since in real space the fluctuations 
are real, one finds $\delta \epsilon^0_{\rm fi}(\vek{q})^* 
= \delta \epsilon^0_{\rm fi}(-\vek{q})$. 
Furthermore  we left out well-known factors that can be found, e.g. in 
the book of Berne and Pecora \cite{Berne1976}.

In the case of longitudinal phonons, the expression for 
$\delta \epsilon_{ij}$ used in \gl{epsQonly} is incomplete. A contribution
proportional to the mass density fluctuation, $\delta \rho$, has to be added,
leading to:
\begin{equation}\label{eps}
\delta \epsilon_{ij}(\vek{r},t) = a 
\delta \rho(\vek{r},t) \delta_{ij} 
+ b Q_{ij}(\vek{r},t) \, ,
\end{equation}
where $\delta \rho$ and $Q_{ij}$ are here the density 
change and its molecular orientation counterpart. Their respective spatial
Fourier transforms couple dielectric fluctuations 
to  longitudinal phonons. 

Since the constitutive equations for the stress tensor and the equation
of motion of the orientation are in the form of integro-differential equations
it is convenient to use the Laplace-Transform (LT), which we use with 
the convention for functions
$f(t)$:
\begin{equation}
LT[ f(t) ](\omega) = i \int_0^\infty dt f(t) \exp(-i \omega t) \, .
\end{equation}

The scattered intensity $I_{\rm fi}(\vek{q},\omega)$ can thus be extracted as
the imaginary part of: 
\begin{equation}\label{chifi}
\chi(\hat{\vek{e}}_{\rm i},\hat{\vek{e}}_{\rm f},\vek{q},\omega) = 
LT[ \langle \delta \epsilon_{\rm fi}(\vek{q},t) 
\delta \epsilon_{\rm fi}^0(\vek{q})^{*} \rangle ](\omega) \, ,
\end{equation}
which defines the light-scattering problem in terms of correlation functions.

\subsection{Expressions for the VV geometry}
In the rest of this paper, we shall make use of the usual Berne and Pecora
axes for light scattering, renamed, in agreement with [I], 
${\vek{\parallel}}$ for the direction of $\vek{q}$, 
${\vek{\perp}}$ for the direction perpendicular to the scattering plane, 
and ${\vek{\perp}}'$ 
for the direction perpendicular to ${\vek{\parallel}}$ and 
${\vek{\perp}}$. 
We are interested here in the usual VV scattering geometry where 
$\hat{\vek{e}}_{\rm i}$ and
$\hat{\vek{e}}_{\rm f}$ are parallel to ${\vek{\perp}}$ so that the r.h.s.
of \gl{epsfi} reduces to $a\delta \rho + b Q_{\perp \perp}$. The expression of 
$\chi(\hat{\vek{e}}_{\rm i},\hat{\vek{e}}_{\rm f},\vek{q},\omega)$ is thus 
completely determined by the knowledge of the Laplace transform
of the four correlation functions: $\langle \delta \rho(\vek{q},t) 
\delta \rho^0(\vek{q})^* \rangle, 
\langle \delta \rho(\vek{q},t)  Q_{\perp\perp}^0(\vek{q})^* \rangle, 
\langle  Q_{\perp\perp}(\vek{q},t) \delta \rho^0(\vek{q})^* \rangle$ and  
$\langle Q_{\perp\perp}(\vek{q},t) Q_{\perp\perp}^0(\vek{q})^* \rangle$, 
that we
shall compute in turn. 

The spatial Fourier transforms of 
Equations (\ref{masscons})
 and (\ref{currcons}) can be grouped into the 
single equation:
\begin{equation}\label{ddotrho}
\delta \ddot{\rho}(\vek{q},t) = q^2 \sigma_{\parallel \, \parallel}(\vek{q},t)
\, .
\end{equation}
Inserting the constitutive equation for $\sigma_{\parallel\parallel}$, 
Eq. (\ref{constitutive}), and performing the Laplace transform
yields with the help of Eqs. (\ref{pressure}) and (\ref{taupar}), see below:
\begin{eqnarray}\label{rhopar}
& & -c^2 q^2 \delta \rho(\vek{q},\omega) 
- q^2 \mu(\omega) \left[ 
\omega Q_{\parallel\parallel}(\vek{q},\omega) - 
Q_{\parallel\parallel}^0(\vek{q}) \right] = \nonumber \\
& & \left[ \frac{q^2}{\rho_m}  [\eta_b(\omega)+\frac{4}{3} \eta_s(\omega)]
 - \omega \right]  
\left[ \omega \delta \rho(\vek{q},\omega) - \delta \rho^0(\vek{q}) \right] 
\, ,
\end{eqnarray}
Here,  we left out 
the term containing $\dot{\rho}^0(\vek{q})$ since it will drop out once 
correlation functions with variables of even time parity are built. 
Due to the coupling we also need 
 the Fourier-Laplace transform of 
\gl{Qosci}:
\begin{eqnarray}\label{LTQ}
&& - \omega_0^2 Q_{ij}(\vek{q},\omega) - i \Lambda' \mu(\omega) 
\tau_{ij}(\vek{q},\omega) = \nonumber \\
& & \left[ \Gamma'(\omega) - \omega \right] 
\left[ \omega Q_{ij}(\vek{q},\omega)  -Q_{ij}^0(\vek{q})  \right]
\, .
\end{eqnarray} 
As above,  the term containing $\dot{Q}_{ij}^0(\vek{q})$ 
has been dropped 
since we will correlate \gl{LTQ} with quantities of even time parity only.
Using: 
\begin{eqnarray}\label{taupar}
\tau_{\parallel\parallel}(\vek{q},\omega) & = &  
-\frac{4}{3} i q v_{\parallel}(\vek{q},\omega) \nonumber \\
& = &  -\frac{4i }{3 \rho_m} \left[ \omega 
\delta \rho(\vek{q},\omega) - \delta \rho^0(\vek{q}) \right]
\, ,
\end{eqnarray} 
\begin{eqnarray}
\tau_{\perp\perp}(\vek{q},\omega) & = &  
\frac{2}{3} i q v_{\parallel}(\vek{q},\omega) \nonumber \\
& = &  \frac{2i }{3 \rho_m} \left[ \omega 
\delta \rho(\vek{q},\omega) - \delta \rho^0(\vek{q}) \right]
\, ,
\end{eqnarray} 
one obtains from Eq. (\ref{LTQ}) for the orientational fluctuations:
\begin{eqnarray}\label{Qpar}
Q_{\parallel\parallel}(\vek{q},\omega) 
& = &   
 - \frac{4\Lambda'}{3 \rho_m} r(\omega) 
\left[
\delta \rho(\vek{q},\omega) 
- \delta \rho^0(\vek{q})/\omega \right]  \nonumber \\
& & 
 + \left( 1 - \frac{\omega_0^2}{D(\omega)} \right) 
\frac{ Q^0_{\parallel\parallel}(\vek{q})}{\omega} \, ,
\end{eqnarray}
\begin{eqnarray}\label{Qperp}
Q_{\perp\perp}(\vek{q},\omega) 
& = &   
  \frac{2\Lambda'}{3 \rho_m} r(\omega) 
\left[
\delta \rho(\vek{q},\omega) 
- \delta \rho^0(\vek{q})/\omega \right]  \nonumber \\
& & 
 + \left( 1 - \frac{\omega_0^2}{D(\omega)} \right) 
\frac{ Q^0_{\perp\perp}(\vek{q})}{\omega} \, .
\end{eqnarray}
We have introduced, here, the quantity $D(\omega)$ which determines
the frequency dependence of the pure orientational motions: 
\begin{subequations}\label{Drdef}
\begin{equation}
D(\omega) = \omega_0^2 + \omega \Gamma'(\omega) - \omega^2 \, ,
\end{equation}
and for  later use we also introduce the rotation-translation coupling: 
\begin{equation}
r(\omega) = \omega \mu(\omega) [D(\omega)]^{-1} \, .
\end{equation}
\end{subequations}
From Eqs. (\ref{rhopar},\ref{Qpar},\ref{Qperp}) 
one can solve for orientational and
density fluctuations in terms of their respective initial values:
\begin{eqnarray}\label{rhoqom}
\omega \delta \rho(\vek{q},\omega) &= &  
 q^2 A(\vek{q},\omega)
+  \delta \rho^0(\vek{q}) \, , 
\end{eqnarray}
\begin{eqnarray}\label{Qompar}
\omega Q_{\parallel \parallel}(\vek{q},\omega) & = & 
\left( 1 - \frac{\omega_0^2}{D(\omega)} \right) 
Q_{\parallel\parallel}^0(\vek{q}) \nonumber \\
& & 
 - \frac{4\Lambda'r(\omega)}{3 \rho_m} 
 q^2 A(\vek{q},\omega)   \, ,
\end{eqnarray}
\begin{eqnarray}\label{Qpperp}
\omega Q_{\perp \perp}(\vek{q},\omega) & = & 
\left( 1 - \frac{\omega_0^2}{D(\omega)} \right) 
Q_{\perp\perp}^0(\vek{q}) \nonumber \\
& & 
 + \frac{2\Lambda'r(\omega)}{3 \rho_m} 
 q^2 A(\vek{q},\omega)   \, ,
\end{eqnarray}
with:
\begin{equation}
A(\vek{q},\omega) = P_L(q,\omega)
\left[ c^2 
\delta \rho^0(\vek{q}) - \omega_0^2 r(\omega) 
Q_{\parallel \parallel}^0(\vek{q}) \right] \, .
\end{equation}
Here $P_L(q,\omega)$ abbreviates the longitudinal phonon
propagator:
\begin{eqnarray}\label{longiPL}
P_L(q,\omega) = \left[ \omega^2 - q^2 c^2 
- q^2 \omega \eta_L(\omega) /\rho_m \right]^{-1} \, ,
\end{eqnarray}
and, in this equation,
the coupling to the parallel component of all the damping 
mechanisms is expressed
in terms of the longitudinal viscosity:
\begin{eqnarray}
\eta_L(\omega) = \eta_b(\omega) + \frac{4}{3} \left[ 
\eta_s(\omega) - \frac{\Lambda'}{\omega} D(\omega) r(\omega)^2 \right] \, .
\end{eqnarray}
The density-density and the 
density-orientation
 correlation functions are then
obtained from Eqs. (\ref{rhoqom},\ref{Qompar}):
\begin{subequations}\label{rhorho}
\begin{eqnarray}
\lefteqn{ LT[ \langle \delta \rho(\vek{q},t) 
\delta \rho^0(\vek{q})^{*} \rangle ](\omega)
=  \nonumber } \\
& &  \frac{1}{\omega} \left[ 1 + q^2 c^2 P_L(q,\omega) \right] 
\langle |\rho^0(\vek{q})|^2 \rangle \, ,
\end{eqnarray}
\begin{eqnarray}\label{rhoperpperp}
& & LT[ \langle \delta \rho(\vek{q},t) Q_{\perp \perp}^0(\vek{q})^{*} 
\rangle ](\vek{q},\omega)
=  
\nonumber \\
& &  -q^2 \frac{\omega_0^2 r(\omega)}{\omega} P_L(q,\omega) 
\langle Q_{\parallel \parallel}^0(\vek{q})^* Q_{\perp \perp}^0(\vek{q}) 
\rangle \, ,  
\end{eqnarray}
\end{subequations}
once one has taken into account that both $\langle | \rho^0(\vek{q})|^2 
\rangle$ and 
$\langle Q_{\parallel \parallel}^0(\vek{q})^* Q_{\perp \perp}^0(\vek{q}) 
\rangle $ are of order unity, while:
\begin{eqnarray}\label{qsquared}
\langle \delta \rho^0(\vek{q})^* Q_{\perp \perp}^0(\vek{q}) \rangle 
= 
\langle Q_{\perp \perp}^0(\vek{q})^* \delta\rho^0(\vek{q}) 
\rangle = {\cal O}(q^2 l^2) \, ,
\end{eqnarray}
where $l$ is a typical intermolecular distance. 
The preceeding relation derives from the fact that one can write, for 
instance:
\begin{eqnarray}
& & \langle Q_{\parallel \parallel}^0(\vek{q})^* \rho^0(\vek{q}) 
\rangle = \nonumber \\ & & 
 \frac{m}{3N} \langle \sum_{\alpha=1}^N \sum_{\beta=1}^N 
e^{i \vek{q} \cdot (\vek{R}_\alpha -\vek{R}_\beta)}
\left[ 3 \hat{{u}}_{\parallel \beta}^2 -1 \right] \rangle \, ,
\end{eqnarray} 
where $\vek{R}_\alpha$ (resp. $\vek{R}_\beta$) are molecular centre of mass
positions and $\hat{u}_{\beta \parallel}$ is the projection of the unit 
vector $\hat{\vek{u}}$ of the molecule $\beta$ on the direction
${\vek{\parallel}}$. The exponential factor in the preceeding 
equation can be expanded in powers of $\vek{q} \cdot (\vek{R}_\alpha 
- \vek{R}_\beta)$. One easily convinces oneself that, for symmetry 
reasons, the first two terms of the expansion average to zero so that the
first nontrivial contribution comes from the $[\vek{q} \cdot (\vek{R}_\alpha 
- \vek{R}_\beta)]^2$ term, which is ${\cal O}(q^2)$. Conversely, 
using the same argument for the averages appearing in Eqs. (\ref{rhorho}), one 
finds that the first 
term of the expansion already gives a nonzero value in the long-wavelength 
limit and is 
proportional to the absolute temperature. The dependence on $\vek{q}$
can thus be omitted in those expressions and will also be, in the 
rest of the paper, for all those equal-time averages. 

For the VV scattering 
one has to study also the auto-correlation of the $\perp\perp$ component 
in the orientation. From Eq. (\ref{Qpperp}) one obtains:
\begin{eqnarray}\label{Qdelrho}
& & LT[\langle Q_{\perp\perp}(\vek{q},t) \delta \rho^0(\vek{q})^{*} \rangle]
(\omega) = \nonumber \\
& & \frac{2\Lambda'}{3\rho_m} \frac{r(\omega)}{\omega} c^2 
q^2 P_L(q,\omega) \langle | \rho^0 |^2 \rangle \, ,
\end{eqnarray}
\begin{eqnarray}\label{Qperp2}
& & LT[\langle Q_{\perp\perp}(\vek{q},t) Q_{\perp\perp}^0(\vek{q})^{*} \rangle
](\omega) = 
\frac{1}{\omega} 
\left( 1-\frac{\omega_0^2}{D(\omega)} \right) 
\langle | Q_{\perp\perp}^0|^2 \rangle 
\nonumber \\
& & 
-  \frac{2\Lambda'}{3\rho_m} \omega_0^2 
 \frac{r(\omega)^2}{\omega} q^2 P_L(q,\omega)
 \langle Q_{\parallel\parallel}^0 Q_{\perp\perp}^0
\rangle 
\, .
\end{eqnarray}
Here some comments are in order.

a) One readily verifies that, would one have taken the $\perp\perp'$ 
component of \gl{LTQ} instead of its $\perp\perp$ component,
one would have obtained:
\begin{eqnarray}\label{IVHomega}
& & LT[\langle Q_{\perp\perp'}(\vek{q},t) Q_{\perp\perp'}^0(\vek{q})^{*} 
\rangle
](\omega) = \nonumber \\
& & \frac{1}{\omega} 
\left( 1-\frac{\omega_0^2}{D(\omega)} \right) 
\langle | Q_{\perp\perp'}^0|^2 \rangle \, .
\end{eqnarray}
Up to a $b^2$ factor, the imaginary part of the r.h.s. of the
preceeding equation is simply what is referred to as the back
scattering depolarised spectrum, $I_{VH}(\omega)$, which is not sensitive
to $q$ for long wavelengths. 
The imaginary part of the first term of \gl{Qperp2} is thus the non-acoustic 
part $I_{VV}(\omega)$, up to the same $b^2$ factor (see \gl{eps}). 
Furthermore, since $Q_{ij}$ is a traceless tensor of order two 
\cite{Berne1976}:
\begin{subequations}\label{tensorsym}
\begin{equation}
\langle | Q_{\perp\perp}^0 |^2 \rangle = \frac{4}{3}
\langle | Q_{\perp\perp'}^0 |^2 \rangle \, ;
\end{equation}
the first term of the r.h.s of \gl{Qperp2} represents the well-known
result that the $q$-independent part of $I_{VV}$ coincides with 
$4/3$ of the back scattering spectrum $I_{VH}(\omega)$.

b) The four other terms, Eqs. (\ref{rhorho},\ref{Qdelrho})
 and the second term of the 
r.h.s of \gl{Qperp2} all contain the phonon propagator, $P_L(q,\omega)$, 
and are thus $q$-dependent. Also, 
due to the spherical symmetry of the liquid:
\begin{eqnarray}\label{spherical}
 \langle Q_{\perp\perp}(\vek{q},t) 
\delta \rho^0(\vek{q})^{*} \rangle
=
\langle \delta \rho(\vek{q},t) 
Q_{\perp\perp}^0(\vek{q})^{*} \rangle \, .
\end{eqnarray}
Comparing Eqs. (\ref{rhoperpperp}) and (\ref{Qdelrho}), 
the preceeding 
relation implies:
\begin{equation}\label{equipartition}
\langle Q_{\parallel\parallel}^0 Q_{\perp\perp}^0 \rangle = 
- \frac{2 \Lambda'}{3\rho_m } \frac{c^2}{\omega_0^2} 
\langle | \rho^0 |^2 \rangle \, ,
\end{equation}
while, because $Q_{ij}$ is a traceless tensor, one also has the 
relationship:
\begin{eqnarray}
\langle Q_{\parallel \parallel}^0 Q_{\perp\perp}^0 \rangle = -\frac{1}{2}
\langle | Q_{\parallel \parallel}^0 |^2 \rangle \, .
\end{eqnarray}
\end{subequations}
Inserting these results into Eqs. (\ref{Ifi},\ref{eps}), 
one finally obtains for the intensity in a VV scattering experiment:
\begin{eqnarray}\label{IVV}
& & I_{VV}(q,\omega)  =  \frac{\langle | Q_{\perp\perp'}^0 |^2 \rangle 
}{\omega} Im \Bigg\{ 
\frac{4b^2}{3} \left[ 1- \frac{\omega_0^2}{D(\omega)} \right] 
\nonumber \\
& & 
+\frac{\rho_m}{\Lambda'} (\omega_0 q)^2 P_L(q,\omega) 
\left[a + \frac{2\Lambda'}{3\rho_m} b r(\omega) \right]^2 \Bigg\}
\, ,
\end{eqnarray} 
where the preceeding formula makes clear that the first term is 
4/3 the $I_{VH}(\omega)$ back scattering spectrum.

The preceeding results call for two remarks.

a) Equation (\ref{equipartition}) simply expresses the equipartition 
of energy between the centre of mass motions and the libration motions:
\begin{equation}\label{equi}
\frac{3\omega_0^2}{4\Lambda'} \langle | Q_{\parallel\, \parallel}^0 |^2 \rangle
= \frac{c^2}{\rho_m} \langle |\rho^0 |^2 \rangle \propto k_B T
\, ,
\end{equation} 
and this is a necessary condition for the consistency of the 
phenomenological theory summarised in Section I. Equation (\ref{equi})
will appear in a natural way in the microscopic derivation of the
phenomenological equations (see Eqs. (27c,28b) of Part II).

b) Equation (\ref{IVV}) completes previous results obtained in 
\cite{Latz2001}, 
separating clearly the role of the phonon propagator, $P_L(q,\omega)$
from that of the scattering mechanisms ($a$ and $b$, with appropriate factors).
It also separates the non-hydrodynamic,  rotational contribution
(first term of \gl{IVV}) from the hydrodynamic one. 
In the latter, one part (term in $b^2$) is entirely
due to scattering by the anisotropic part of the polarisability tensor of the 
molecules. The same mechanism was at the origin of the $q$-dependent
part of $I_{VH}(q,\omega)$ in [I]  and
in both cases the scattered intensity is proportional to a phonon
propagator, longitudinal or transverse, multiplied by the same $r(\omega)^2$ 
factor; the power two stresses that the molecular orientation acts twice,
once as the source of the fluctuation and, the second time, as
 the detection mechanism. The more complex form obtained here 
for the $q$-dependent part of $I_{VV}(q,\omega)$ is the 
direct consequence of the existence of two parallel channels, $\delta \rho$ and
$Q_{ij}$, in the case of longitudinal phonons; this is in contradistinction
with the
single channel case of the transverse phonons [I].

\subsection{Results for the HH geoemtry}
The same technique can be used to derive the intensity which can be obtained in
an HH experiment; indeed in such a case:
\begin{eqnarray}
\delta \epsilon_{HH}(\vek{q},t) = \delta \epsilon_{\perp'\perp'}(\vek{q},t)
\sin^2 \frac{\theta}{2} -
\delta \epsilon_{\parallel\parallel}(\vek{q},t) \cos^2 \frac{\theta}{2} 
\, ,
\end{eqnarray}
where $\theta$ is the scattering angle. Expanding in terms of density and
orientation, one has to consider: 
\begin{eqnarray}\label{epsHH}
\delta \epsilon_{HH}(\vek{q},t) & = & - a 
\delta \rho(\vek{q},t) \cos \theta 
+ bQ_{\perp'\perp'}(\vek{q},t) \sin^2 \frac{\theta}{2} 
\nonumber \\ & & 
- bQ_{\parallel\parallel}(\vek{q},t) \cos^2 \frac{\theta}{2} \, .
\end{eqnarray}
To calculate the corresponding auto-correlation function as required 
by Eq. (\ref{Ifi}), one needs again the dynamics of density and orientation. 
Repeating the calculations of Sec. II.A, one convinces oneself that Eq.
(\ref{Qpperp}) remains valid if 
$\perp\perp$ is replaced by $\perp'\perp'$.  
Due to rotational symmetry, the cross correlators are again identical as 
in Eq. (\ref{spherical}). Of the six correlation functions that can be 
built in Eq. (\ref{epsHH}), one can easily compute the missing three 
by the methods of the preceeding subsection. From Eq. (\ref{Qompar}), 
one finds: 
\begin{eqnarray}
& & 
LT[ \langle Q_{\parallel\parallel}(\vek{q},t) Q_{\perp'\perp'}^0(\vek{q})^* 
\rangle ](\omega)  =  \Bigg\{  \left( 
1 - \frac{\omega_0^2}{D(\omega)} \right) \nonumber \\
& &  
+ \frac{4 \Lambda' \omega_0^2}{3 \rho_m} r(\omega)^2 q^2
 P_L(\vek{q},\omega) \Bigg\} 
\frac{\langle Q_{\parallel\parallel}^0 Q_{\perp'\perp'}^0 \rangle}{\omega} 
\, ,
\end{eqnarray}
\begin{eqnarray}
& & 
LT[ \langle Q_{\parallel\parallel}(\vek{q},t) 
Q_{\parallel\parallel}^0(\vek{q})^* 
\rangle ](\omega)  =  \Bigg\{  \left( 
1 - \frac{\omega_0^2}{D(\omega)} \right) \nonumber \\
& &  
+ \frac{4 \Lambda' \omega_0^2}{3 \rho_m} r(\omega)^2 q^2
 P_L(\vek{q},\omega) \Bigg\} 
\frac{\langle | Q_{\parallel\parallel}^0 |^2 \rangle}{\omega} 
\, ,
\end{eqnarray}
\begin{eqnarray}
& & 
LT[ \langle Q_{\parallel\parallel}(\vek{q},t) 
\delta \rho^0(\vek{q})^* 
\rangle ](\omega)  = \nonumber \\
& &  -
 \frac{4 \Lambda' c^2}{3 \rho_m} r(\omega) q^2
 P_L(\vek{q},\omega)  
\frac{\langle | \delta \rho^0 |^2 \rangle}{\omega} 
\, ,
\end{eqnarray}
Collecting all the terms appearing in Eqs. (\ref{chifi},\ref{epsHH}) 
with $\delta \epsilon_{\rm fi} = \delta \epsilon_{HH}$, one arrives at our 
final expression for the HH scattering spectrum:
\begin{eqnarray}\label{IHH}
& & I_{HH}(q,\omega) = \frac{\langle | Q_{\perp\perp'}^0 |^2 \rangle }{\omega} 
\Bigg\{ \frac{4b^2}{3} \left( 1 - \frac{1}{4} \sin^2 \theta \right) 
\left[ 1- \frac{\omega_0^2}{D(\omega)} \right] \nonumber \\
& & + 
\frac{\rho_m \omega_0^2}{\Lambda'} 
 q^2 P_L(q,\omega) 
\left[ a \cos \theta -\frac{b \Lambda' r(\omega)}{3 \rho_m}  (3 +\cos \theta) 
 \right]^2
\Bigg\} 
\, .
\end{eqnarray}

For a given $\vek{q}$, all other scattering geometries give intensities, 
which are linear combinations of $I_{VV}(q,\omega)$ and 
$I_{HH}(q,\omega)$  as well as of $I_{VH}(q,\omega)$, whose
$q$-dependent part detects the transverse phonon, whereas the $q$-independent
part detects again $I_{VH}(\omega)$, Eq. (\ref{IVHomega}). It
is thus not possible to detect independently the contributions of the 
three terms entering, e.g. \gl{IVV}. 
Equations (\ref{IVV}) and (\ref{IHH}) are very compact expressions in which 
the $q$-dependent and the $q$-independent parts have been separated. 
They allow to study, within some approximations for the different
quantities entering them, the influence of the
rotation-translation coupling on the shape of a typical longitudinal 
phonon spectrum. Using for those quantities the same toy model as 
in \cite{Latz2001}, we shall proceed to this comparison in the next
Section.

\section{Analytical and numerical discussion}

\subsection{Introduction}
In this Section, we discuss some typical features of the spectra 
obtained in a VV scattering experiment. As we shall see, the 
rotation-translation coupling gives rise to spectra which differ 
in shape and, more importantly, in some cases 
 in the values for the 
relaxation times, would they be extracted 
 from the widely used model where  density 
fluctuations are considered as the sole light scattering mechanism  
 for the $q$-dependent part of the polarized spectrum, i.e. : 
\begin{eqnarray}\label{single} 
 I_{VV}^{single}(q,\omega)  & = &   \frac{4}{3} I_{VH}(\omega) 
\nonumber \\ 
& & + 
\frac{a^2  c^2 q^2}{\omega} \langle | \rho^0 |^2 \rangle 
Im  P_L(q,\omega) 
 \, .
\end{eqnarray} 

In order to demonstrate the effects which take place 
when  translation-rotation coupling 
is non-negligible, we generate data for the $q$-dependent part of 
a VV experiment by using 
the full expression, \gl{IVV}. The spectral shape is then determined by 
the frequency dependence of the four fundamental memory kernels 
$\eta_s(\omega), \eta_b(\omega), \mu(\omega)$ and $\Gamma'(\omega)$. 
To keep the discussion as simple as possible,  we shall
model the relaxation kernels by single exponential Debye processes:  
\begin{equation}\label{Debye}
M(\omega) = \Delta_M^2 \frac{i \tau}{1+ i \omega \tau} \, ,
\end{equation}
where $M$ is any of $\eta_s, \eta_b, \mu$ or $\Gamma'$. Once the spectra
have been generated, we treat them as raw data and analyse them 
by a 'density-only' model, \gl{single}, in which, see Eq. (\ref{longiPL}),
$\omega_L = c q$ and $\eta_L(\omega)$ have to be fitted; 
in the spirit of the simplifying assumptions made 
above, $\eta_L(\omega)$ is 
also a simple Debye process,  characterised  as in \gl{Debye}, 
 by an 
amplitude, $\Delta_L^2$, and  a fitted relaxation time, $\tau_L$.

Let us first 
make some comments on the general shape of the spectra. For 
a given wave vector $\vek{q}$, the second part of the r.h.s of 
Eqs. (\ref{IVV} or \ref{IHH}) is the sum of three terms 
that correspond to three channels of detection of the longitudinal
phonons: a density-only channel, proportional to $a^2$, an
orientation-only one ($\propto b^2$) and a
cross channel of density and orientation ($\propto a b$).
For a simple liquid, the first one is predominant, which depends solely on
$P_L(q,\omega)$ and the shape of the corresponding spectrum has been
studied in detail, in particular for supercooled liquids, in many papers 
\cite{Dreyfus1992,Du1994,Aouadi2000,Shen2000,Monaco2001}. 
The purpose of these papers was basically to extract information 
on the memory function $\eta_L(\omega)$, and on its  dependence on 
temperature. In particular, if $\eta_L(\omega)$ is characterised 
by a single Debye relaxation process with a relaxation 
time $\tau_L$ and an amplitude such that, whatever $\tau_L$, 
$Im P_L(q,\omega)/\omega $ exhibits a well-defined peak,  
 the density spectrum varies in a specific manner 
with $\tau_L$: for $\omega_B \tau_L \ll 1$, with  $\omega_B = c q$,  
the spectrum
 contains a quite narrow Brillouin peak centered around, say $\omega_{B1}$,
with a weak background for other frequencies. 
The same is true for $\omega_B \tau_L \gg 1$, but the peak is now 
centered at $\omega_{B2} \gg \omega_{B1}$. The only regime where a 
rather precise information on $\tau_L$ is obtained occurs for 
$\omega_B \tau_L \sim 1$, i.e. approximately $1/5 \lessapprox \,  \, 
\omega_B \tau_L 
\lessapprox \, \, 5$. 

The influence of the orientation-only channel on the shape of the
$q$-dependent part of the VV spectrum was discussed in \cite{Latz2001},
with the help of a similarly simple model for the four relaxation kernels; 
conversely,
the cross channel ($\propto ab$) was not studied in that paper. 
We will thus 
first compare the shape of the contributions of the three channels. 
We shall then discuss the line-shape of the total $q$-dependent part 
of the VV spectrum, focusing mostly on the $\omega_B \tau \sim 1$ regime. 
To remain as close as possible to the results presented 
in \cite{Latz2001}, we take the same numerical values
for the various temperature independent parameters, $c =0.6, q=0.02,  
\omega_0 = 1$, the units being chosen such that $\Lambda' = 1$ and 
$\rho_m = 1$. In order to restrict the number of parameters, 
the relaxation time $\tau$ is the same for all four 
memory kernels indicated in \gl{Debye}, and we use the same values of $\tau$
as 
in \cite{Latz2001},
 simply studying also and in more details the $\omega_B \tau
\sim 1$ regime.  The amplitudes are taken as $\Delta_{\eta_b} = 1/\sqrt{2}, 
\Delta_{\eta_s} = \sqrt{3/8}$ and $\Delta_{\mu} = \sqrt[4]{3/16}$ in order to
use the same phonon propagator as in \cite{Latz2001}.
Since the Brillouin peak is at much smaller frequencies than
the libration frequency, $\omega_0$, the spectra are not 
sensitive to the value of $\Delta_{\Gamma'}$ in the frequency window studied
here. Hence, we set $\Delta_{\Gamma'} = 0$, or $\Gamma'(\omega) \equiv 0$. 
Finally, one needs a plausible numerical value for the ratio 
$2\Lambda' a/3 \rho_m b$. Its estimate can be obtained from the following 
considerations. Impulsive Stimulated Thermal Scattering 
(ISTS) experiments have been performed on supercooled 
liquids formed of anisotropic molecules such as salol and OTP. 
It has been recognised recently that the spectral shape of the 
signal obtained in this pump-probe
experiment is sensitive to the polarisation of the probe beam 
\cite{Tashin2001,Glorieux2002}.
In
these experiments, the origin of the signal is not the thermal fluctuations
of $\delta \rho$ and $Q_{ij}$  but the 
intensity of the pump beam; conversely, the signal is detected through the 
dependence of the local dielectric tensor on density and orientation 
as in a VV experiment. The ISTS signal
then turns out to 
be linear in 
$\delta \rho$ and $Q_{ij}$ with the same coefficients as in 
\gl{IVV}.  
Making use of different polarisations of the probe 
beam, the contribution of the two variables can be disentangled 
\cite{Tashin2001},
and a value for a typical molecular liquid,
metatoluidine, of the order of $0.5$ can be derived from the results of
\cite{Tashin2001}. In order to overemphasise the effect of the 
additional scattering channels, we have found it convenient,
in the present paper, to choose a value equal to $4/\sqrt{27} \approx 0.77$.
All the results presented in this Section have been obtained with those
numerical values.

\subsection{Analysis of the role of the two additional
channels}

\subsubsection{The $\omega_B \tau \ll 1$ and the $\omega_B \tau \gg 1$
regimes}

\begin{itemize}
\item Preliminary remarks 
\end{itemize}
As already alluded in part A of this Section, 
the two regimes $\omega_B \tau \ll 1$
and $\omega_B \tau \gg 1$ are rather uninteresting from the practical point of 
view of determining a value of $\tau$: the most important 
piece of information is contained in the position of the Brillouin peak,
its linewidth being much more difficult to analyse because its exact 
line-shape cannot be determined. The remaining low intensity part of the 
Brillouin line 
cannot be studied experimentally because it cannot be disentangled,
either from the wing of the central peak (first term of the r.h.s 
of \gl{IVV}) for high temperatures, i.e. $\omega_B \tau \ll 1$, 
or from a flat background for low temperature, $\omega_B \tau \gg 1$. 
Let us just mention here that, even in these regimes, at very
low frequency, the r.h.s of the second term of \gl{IVV} can lead to a
negative contribution to the intensity, due to the existence of the
$r(\omega)$ and $r(\omega)^2$ terms, as was already mentioned 
for the latter case in \cite{Latz2001}. This is not an artefact
of the theory, Eqs. (\ref{constitutive},\ref{Qosci},\ref{eps}), but  
a consequence of the introduction of translation-rotation coupling
into these equations. If one takes it into account, it is no longer
necessary that the  $q$-dependent part of \gl{IVV} is always positive,  
the requirement being that only the full r.h.s of this equation is
positive. We shall differ the discussion of this point to Section IV and,
mostly, to Part II \cite{Franosch2002}.

In spite of the preceeding remark on the $\delta$-like shape of those spectra
in the two regimes, their study turns out to be instructive, because the way 
the two additional channels contribute to the 
spectrum will be similar in the $\omega_B \tau \sim 1$ regime we shall
study later; there,
one will directly detect their influence on the line-shape while their
analytical study is easier in the extreme cases we consider here. 
We shall first develop analytic formulae, then demonstrate that
these effects do show up in the 
line-shape of the $Im [ r(\omega) P_L(q,\omega)/\omega]$ and 
$Im [ r(\omega)^2 P_L(q,\omega)/\omega]$ spectra, if analysed with 
sufficiently large accuracy, and finally show that, nevertheless, the 
$q$-dependent part of the VV spectrum is little affected by these effects
in the  two cases, $\omega_B \tau \ll 1$ and $\omega_B \tau \gg 1$, 
in the region of the Brillouin peak.

\begin{itemize}
\item Analytical results
\end{itemize}

\begin{table*}
\begin{ruledtabular}
\begin{tabular}{|c|ccc|ccc|}
& & $\omega_B \tau \ll 1$ &  
& &  $\omega_B \tau \gg 1$ & \\ 
%& & & & & & \\ 
 & $\omega \ll \omega_B$ & Brillouin peak \, \, & $\omega \gg \omega_B$ &
 $\omega \ll \omega_B$ & Brillouin peak \, \,  & 
$\omega \gg \omega_B \quad $ \\ 
%& & & & & & \\ 
\hline\hline
%& & & & & & \\
$a^2 $ & $\frac{\displaystyle \Delta^2 \tau}{\displaystyle q^2 c^4}$ & 
$\frac{\displaystyle 1}{\displaystyle \Delta^2 q^3 c \tau}$ 
& $\frac{\displaystyle q^2 \Delta^2 \tau}{\displaystyle \omega^4} $ & 
$ \frac{\displaystyle \Delta^2}{\displaystyle \omega^2 
\tau^2 q^2 (c^2 + \Delta^2)^2} $ &
$\frac{\displaystyle \tau}{\displaystyle q^2 \Delta^2}$ & 
$ \frac{\displaystyle q^2 \Delta^2}{\displaystyle \omega^6 \tau} $ \\ 
%& & & & & &  \\
\hline\hline 
%& & & & & & \\
$\quad b^2$ RI $\quad$ & 
$-(\omega \tau)^2$ & & $-(\omega \tau)^2$ & 1 & & 1 \\ 
%& & & & & & \\ 
\hline
%& & & & & & \\
$\quad b^2$ IR $\quad$ & 
$-(\omega \tau)^2 \frac{\displaystyle c^2}{\displaystyle \Delta^2}$
 & & $ (\omega \tau)^2 
\frac{\displaystyle \omega^2}{\displaystyle q^2 \Delta^2}\quad $ & 
$-\frac{\displaystyle c^2 + \Delta^2}{\displaystyle \Delta^2}$ & & 
$\frac{\displaystyle \omega^2}{\displaystyle q^2 \Delta^2}$ \\ 
%& & & & & & \\
\hline\hline
%& & & & & & \\
$\quad ab $ RI $\quad$ 
& $-(\omega \tau)^2$ & & $-(\omega \tau)^2$ & 1 & & 1 \\ 
%& & & & & & \\
\hline  
%& & & & & &   \\
$\quad a b$ IR $\quad $ & $-  \frac{\displaystyle c^2}{\displaystyle \Delta^2}$
 & & $  
\frac{\displaystyle \omega^2}{\displaystyle q^2 \Delta^2}$ & 
$-\frac{\displaystyle c^2 + \Delta^2}{\displaystyle \Delta^2}$ & & 
$\frac{\displaystyle \omega^2}{\displaystyle q^2 \Delta^2}$ \\ 
%& & & & & & \\
\end{tabular}
\end{ruledtabular}
\caption{{\bf Analytic form for the different scattering channels in the two
limiting cases}. The first line, $a^2$, gives the analytic form of 
$Im [P_L(q,\omega)/\omega]$ for $\omega \ll \omega_B$, at the
peak value and for $\omega\gg \omega_B$, for the two cases 
$\omega_B \tau \ll 1$ and $\omega_B \tau \gg 1$. The last four lines
give the factors by which the first line has to be 
multiplied to obtain the analytical form of the corresponding term; the 
factors are given for the $b^2$ channel (second and third lines) and for the 
$ab$ channel (fourth and fifth lines). In both cases, the two 
terms, R(eal) I(maginary) and I(maginary)R(eal), are given in 
the even, respectively odd, lines (see \gl{ImRe}). }
\end{table*}

Table I displays the analytical form of the leading terms for the 
three channels, viz. pure density ($a^2$), pure rotation ($b^2$) 
and cross channel
($ab$), both for $\omega \ll \omega_B$ and $\omega \gg \omega_B$; for
the sake of completeness, the Table also displays the approximate Brillouin peak
intensity related to the first channel. In order to make this
Table as easy to read as possible, the quantities for the $b^2$ and 
$ab$ channels are the factors by which the results of the first line 
($a^2$ channel) have to be multiplied in order to obtain the
corresponding contribution to the intensity. 

Furthermore, while the intensity of the pure density channel is simply equal 
to $Im [ P_L(q,\omega)/\omega]$, in the case of the 
pure orientation channel for instance, its intensity $I_{b^2}(\omega)$ 
is given by: 
\begin{eqnarray}\label{ImRe}
\frac{1}{\omega} Im [  r(\omega)^2
 P_L(q,\omega)] & & =   \frac{1}{\omega} \Bigg\{
Re[ r(\omega)^2 
] Im [P_L(q,\omega)] + 
 \nonumber \\ 
& & 
Im[ r(\omega)^2 ] Re [P_L(q,\omega)] \Bigg\}  \, ,
\end{eqnarray}
and a similar expression holds for the cross channel. Since 
$Im[P_L(q,\omega)/\omega]$ has typical functional form $[1+X^2]^{-1}$
with $X = \omega- \omega_B$ for $\omega>0$, the corresponding real
part $Re[P_L(q,\omega)/\omega]$ is  described by $X/[1+X^2]$; 
the contributions of the two terms   of \gl{ImRe}, and similar terms
in the $ab$ channel, have different line-shapes and have to 
be considered separately. They are denoted as $RI$ and $IR$ 
contributions in the lower part of Table I. Finally, we simplified
our study by admitting that, in the limited frequency range for which this 
Table is constructed,
 $\omega_B/10 \lessapprox \omega \lessapprox 10 \omega_B$,
the dimensionless quantities $\omega\tau$ and $\omega_B \tau$
have always the same order of magnitude. Table I shows that one 
has to study three different cases:

\begin{itemize}
\item[a)] $\omega_B \tau \gg 1$ \\
Within the approximation just mentioned, the $b^2$ and the $ab$ channels 
have the same multiplying factors, and their $RI$ parts 
give rise to exactly the same line-shape as the $a^2$ channel. For 
$\omega \gg \omega_B$, the $(\omega/q\Delta)^2$ factor of the IR term 
increases the intensity with respect to the $a^2$ channel; conversely,
for $\omega \ll \omega_B$, the same IR  term gives a negative 
contribution: the relative intensities will be lower than in the $a^2$ channel.

\item[b)] $\omega_B \tau \ll 1$ \\
-- The $b^2$ channel has a general $(\omega \tau)^2$ factor which 
renders its contribution small on the whole frequency domain whatever its
sign. The IR part gives a negative contribution for $\omega < \omega_B$ which
changes into a positive one at some value above $\omega_B$, while the 
RI part always remains negative. \\ \\
-- The $ab$ channel gives a line-shape quite different from the ones
discussed above since it is dominated by its IR part with a 
prefactor of order unity while the RI part is suppressed by a factor of 
$(\omega \tau)^2$. This results into a typical $X/[1+X^2]$ line shape 
with a negative contribution for $\omega < \omega_B$, a minimum 
below $\omega_B$, a zero in the vicinity of $\omega_B$, and a positive 
contribution above $\omega_B$, the maximum of the latter 
being slightly above $\omega_B$. 
\end{itemize}

\begin{itemize}
\item Numerical results 
\end{itemize}
\begin{figure}[ht!]
\epsfig{file=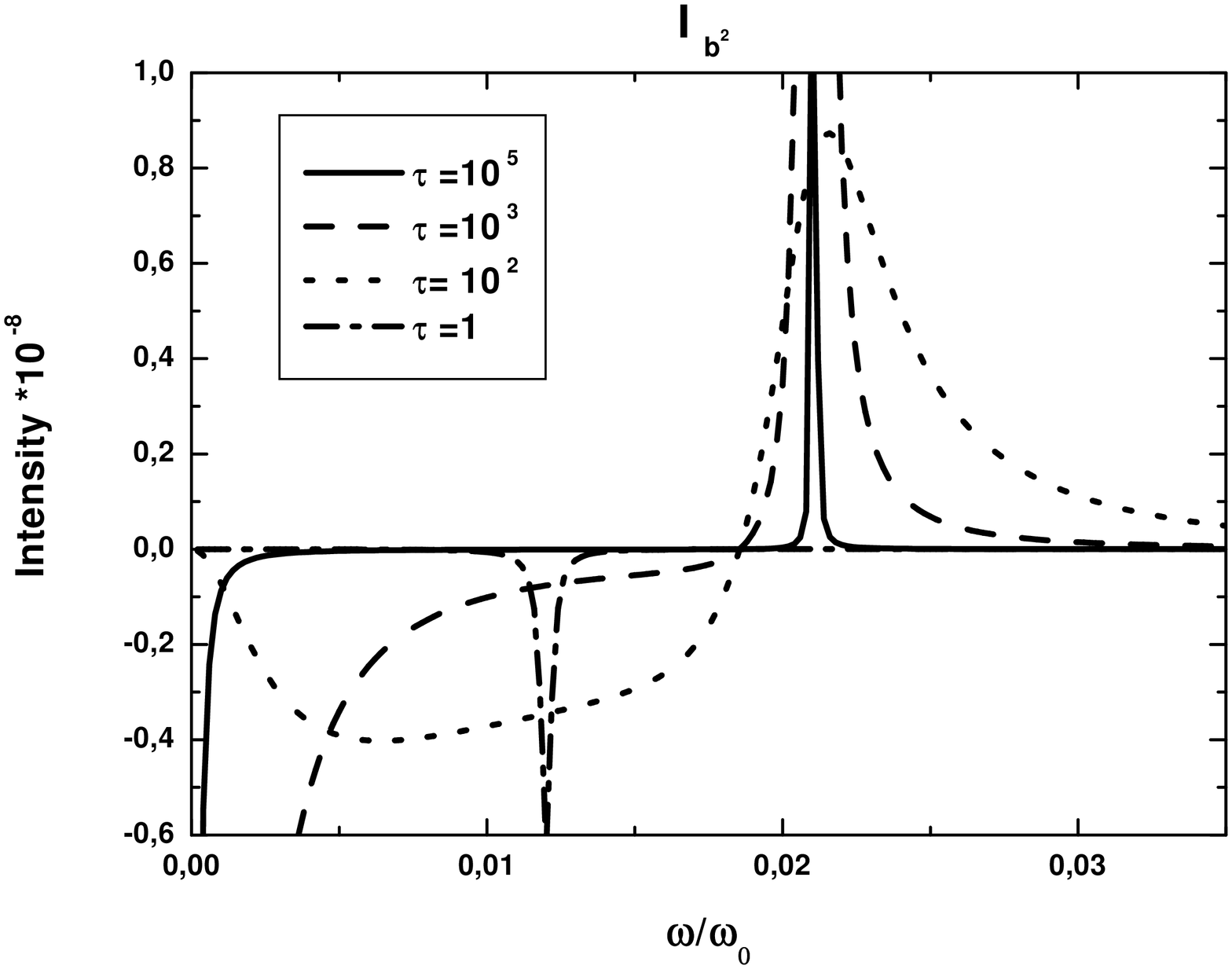,width=7.0 cm,angle=270}
\caption{The intensity of the orientation-only channel $I_{b^2}$ 
for four values of 
$\tau$: $\tau = 1, 10^2, 10^3$ and $10^5$. The spectrum for $\tau = 1$ has 
been multiplied by $100$ to make it visible on the same figure.}
\end{figure}

\begin{figure}[ht!]
\epsfig{file=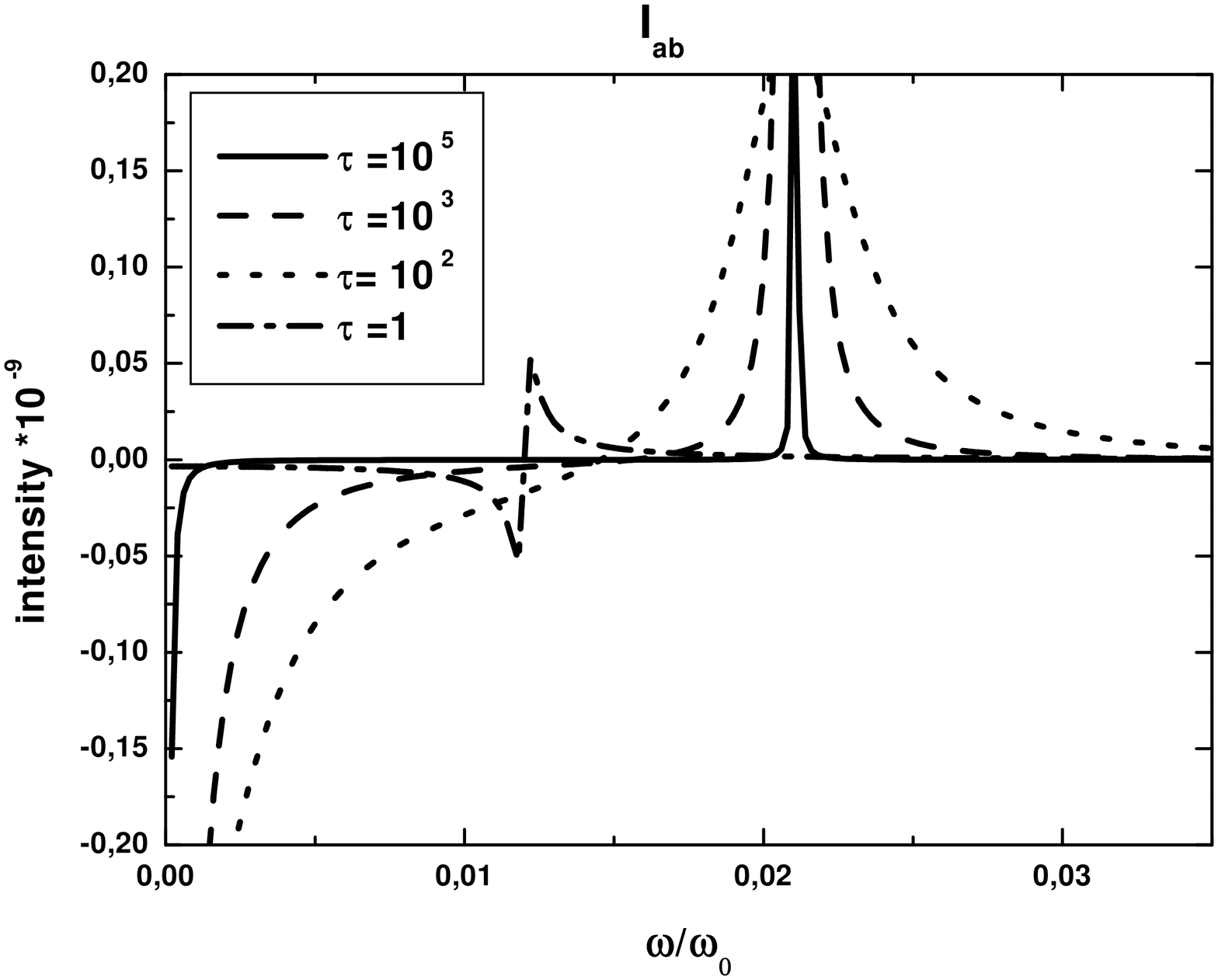,width=7.0 cm,angle=270}
\caption{The intensity of the cross channel $I_{ab}$ for 
the same values of $\tau$ 
as in Fig 1. The shape of the $\tau=1$ spectrum differs strongly from the 
remaining ones.
}
\end{figure}

\begin{figure}[ht!]
\epsfig{file=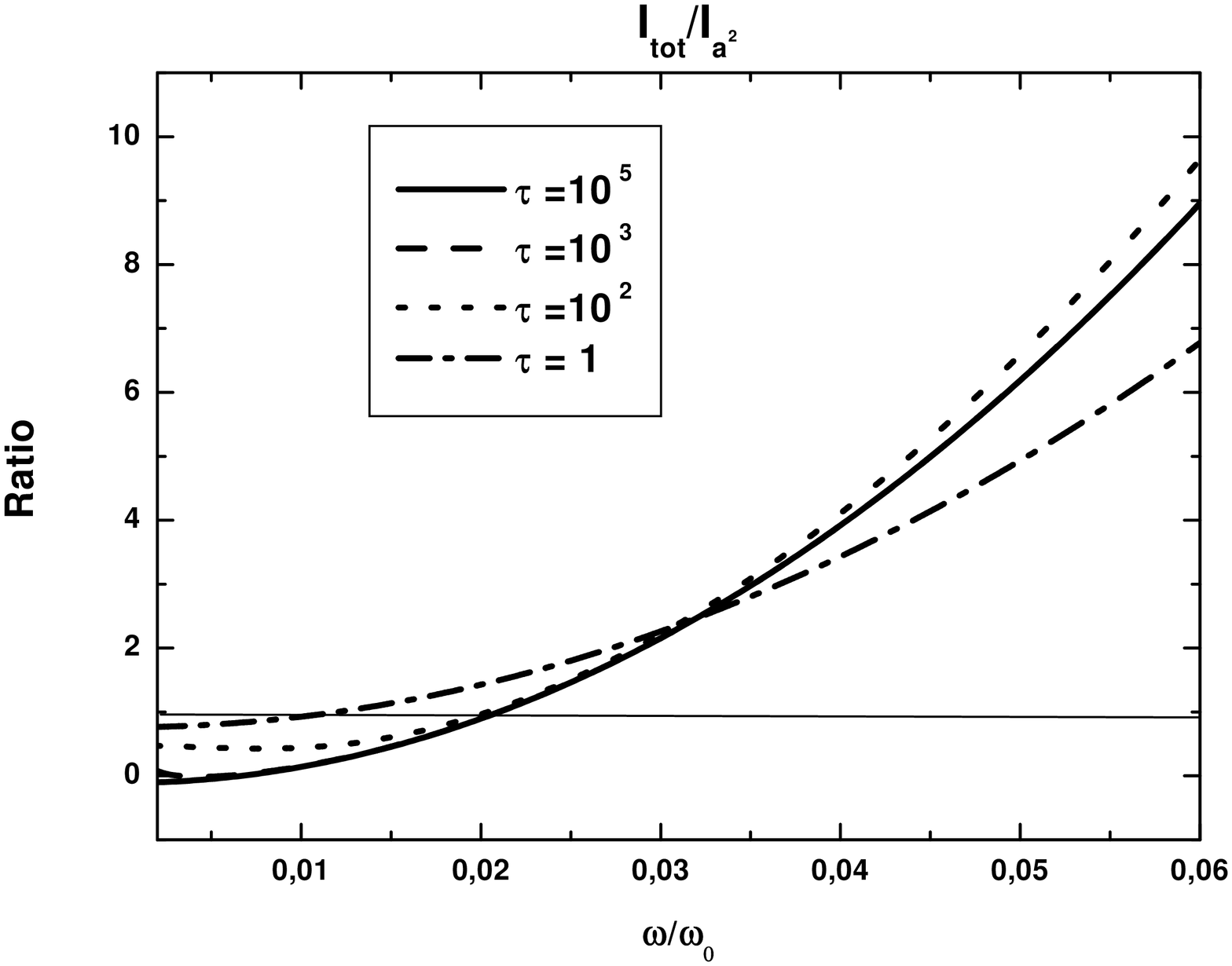,width=7.0 cm,angle=270}
\caption{The ratio of $K I_{tot}$ to $I_{a^2}$ for 
the same values of $\tau$ as in Fig.1. For each spectrum, $K$ is 
adjusted in a such a way that the 
intensities of the two spectra are identical
at the Brillouin peak. The horizontal
line, drawn for a ratio equal to 1, allows to locate these Brillouin peaks 
for the four cases.  
}
\end{figure}

Figures 1,2 and 3 illustrate the preceeding results. Figure 1
displays
the intensity related to $b^2$ channel, $I_{b^2}$,  for $\tau =1, 10^2, 10^3$ 
and $10^5$. The crossover regime  $\omega_B \tau \sim 1$ corresponds
to $\tau \sim 10^2$. The different curves exhibit the shapes 
discussed above. Note that for clarity 
the spectrum corresponding to $\tau = 1$ has
been enhanced by a factor of 100. Yet, the spectra for $\omega_B \tau \gg 1$
exhibit an additional, strongly negative feature for $\omega \ll 
\omega_B$, which was not studied in Table I, as it is related to the 
$\omega \tau \ll 1$ part of these spectra.

The different shapes of the $I_{ab}$ spectra are shown in Fig. 2.
These shapes agree with what can be inferred, in the $\omega_B \tau \gg 1$ 
and $\omega_B \tau \ll 1$ regimes, from the results shown in Table I.

Finally, Fig. 3 represents the ratio of the total $q$-dependent part 
of $I_{VV}$, \gl{IVV}, called $I_{tot}$, 
to the $a^2$ channel intensity alone, $I_{a^2}$. Here the two 
spectra are normalised to the same intensity at the Brillouin peak. 
The factor $(\omega/q\Delta)^2$, general to the IR part 
for $\omega \gg \omega_B$, explains the increase of the ratio 
in this frequency domain, a signature that will remain also in the 
$\omega_B \tau \sim 1$ regime. Also, this ratio always decreases,
more or less strongly, below $\omega_B$ due to the 
negative value of the same term in this frequency range. This effect also
shows up in the $\omega_B \tau \sim 1$ regime, as can be inferred from 
Fig. 3 for the curve corresponding to $\tau = 10^2$.    
Conversely, one observes that the ratio is close to unity 
for frequencies in the vicinity of $\omega_B$. 
 As long as $\omega_B \tau \gg 1$ or
$\omega_B \tau \ll 1$, the shape of the Brillouin line of the total 
$q$-dependent part of $I_{VV}$ is indistinguishable from that of the
pure density channel. Hence, one needs to analyse the $\omega_B \tau \sim 1$ 
regime to detect a difference between the 
'density-only' and the 'density+orientation' scattering model.

\subsubsection{Numerical analysis of the $\omega_B \tau \sim 1$ regime}

\begin{figure*}[ht!]
\epsfig{file=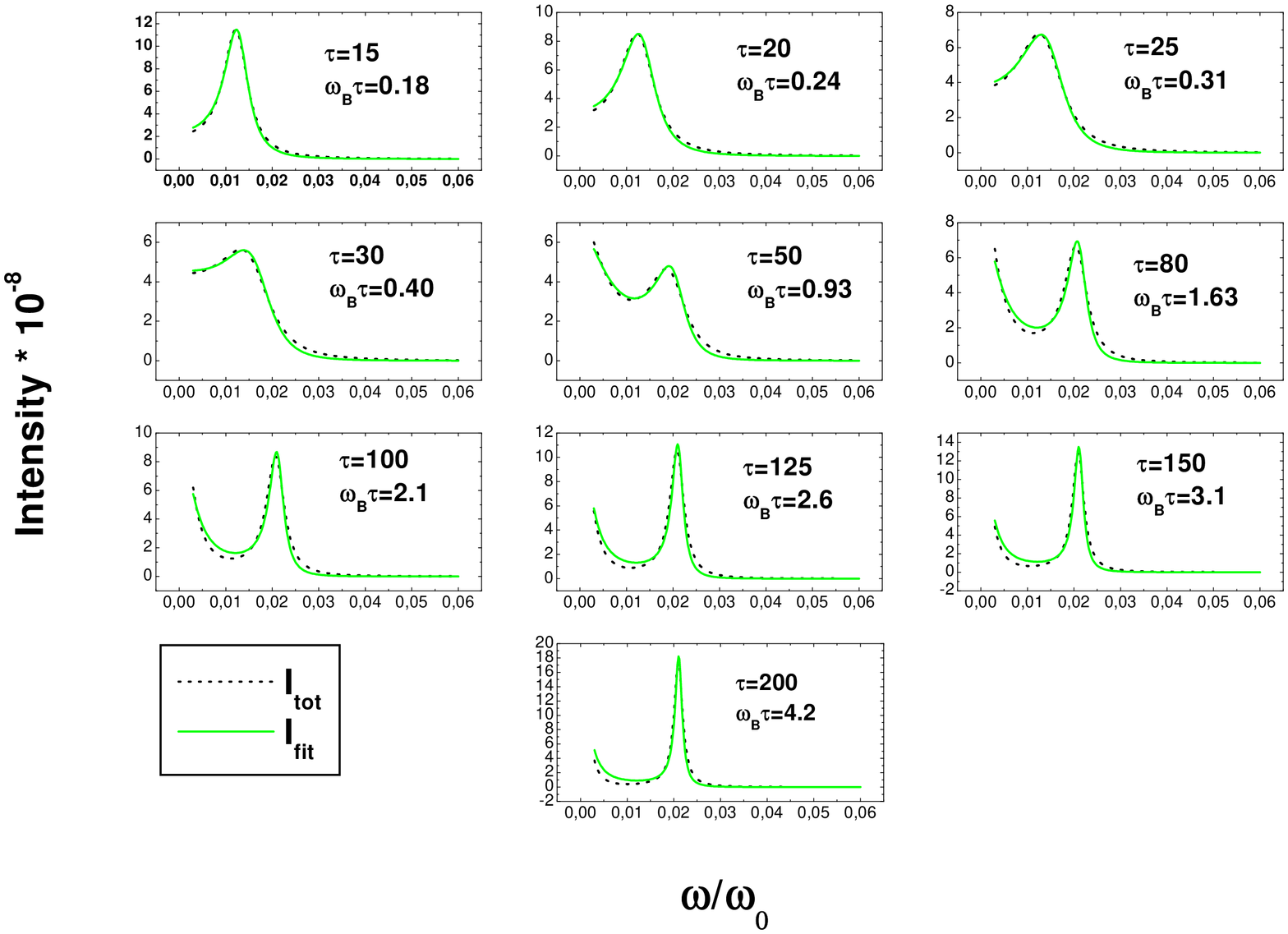,width=14.0 cm,angle=270}
\caption{
$I_{tot}$ (dashed line) for 10 values of $\tau$ in the
vicinity of $\omega_B \tau = 1$, and its fit (full line) with 
a $I_{a^2}$ model (see text). The values of $\tau$ and $\omega_B \tau$ are 
indicated in each case.
}
\end{figure*}
In order to look for an experimentally detectable effect, we 
have computed the total $q$-dependent part of the intensity, $I_{tot}$, 
 for 
the 10 values of $\tau$ reported in the first column of Table II, all
close to the $\omega_B \tau \sim 1$ condition. Those 10 spectra
are shown in Fig. 4 as dashed lines and we 
have checked that they notably differ from the scaled spectra that
can be obtained for the same values of $\tau$
considering a line-shape  as given by \gl{single}. To explore more closely
the effect of the additional channels,
both on the line shape and on a possible misinterpretation of these
spectra in terms of 'density-only' spectra, we have tried to fit them 
with a $Im[P_L(q,\omega)/\omega]$ expression, where the 
longitudinal viscosity was expressed as:
\begin{eqnarray}\label{q2eta}
 q^2  \omega \eta_L(\omega) /\rho_m 
= \Delta_L^2 \frac{i\omega \tau_L}{1+i\omega \tau_L} + i \omega \gamma_L
\, .
\end{eqnarray}
A value of $\gamma_L$, consistent with a previous Brillouin scattering
study of {\it m}-toluidine \cite{Aouadi2000},
 $\gamma_L = 10^{-5}$, was chosen, and
fits for the spectra corresponding to the ten different 
values of $\tau$ were performed, treating $\Delta_L, \omega_L = c q$ and
$\tau_L$ as free parameters.

\begin{table}
\begin{ruledtabular}
\begin{tabular}{c|c|c|c|c}
$\quad\tau\quad$\footnotemark[1] 
 & $\tau_L\quad $\footnotemark[2]  & 
$\Delta_L \quad$\footnotemark[2]  & 
$\omega_L\quad $ \footnotemark[2] & 
$ \omega_B \quad$\footnotemark[3]  \\ 
\hline
15 & $13.4 \quad$ & $0.0210 \quad$ & $0.0125 \quad$ & $0.0121\quad$ \\
20 & $19.8 \quad$ & $0.0200 \quad$ & $0.0128 \quad$ & $0.0123\quad$  \\
25 & $27.6 \quad$ & $0.0190 \quad$ & $0.0131 \quad$ & $0.0125\quad$  \\
30 & $37.0 \quad$ & $0.0180 \quad$ & $0.0133 \quad$ & $0.0132\quad$  \\
50 & $67.7 \quad$ & $0.0170 \quad$ & $0.0141 \quad$ & $0.0186\quad$  \\
80 & $109 \quad$ & $0.0160 \quad$ & $0.0147 \quad$ & $0.0204\quad$  \\
100 & $136 \quad$ & $0.0156 \quad$ & $0.0149 \quad$ & $0.0207\quad$  \\
125 & $175 \quad$ & $0.0154 \quad$ & $0.0148 \quad$ & $0.0208\quad$  \\
150 & $209 \quad$ & $0.0152 \quad$ & $0.0149 \quad$ & $0.0210\quad$  \\
200 & $270 \quad$ & $0.0150 \quad$ & $0.0150 \quad$ & $0.0210\quad$  \\
\end{tabular}
\end{ruledtabular}
\footnotetext[1]{initial value}
\footnotetext[1]{fitted value}
\footnotetext[3]{deduced from $I_{tot}(\omega)$}
\caption{ {\bf 
Fit parameters of the total spectrum for 10 different values of $\tau$}. 
The first column gives the values of $\tau$ 
for which the total spectra, $I_{tot}$,
 are computed. 
The three next columns give the fit values for $\tau_L, \Delta_L$ 
(see \gl{q2eta})
and $\omega_L$, the relaxed phonon frequency, while the last one gives the 
frequency, $\omega_B$, of the peak of $I_{tot}$.
Note that, for these values of $\tau$, $I_{tot}$ exhibits a broad enough maximum (see Fig. 4) 
for $\omega_B$ to be defined with an uncertainty of order $5\times 10^{-4}$.  }
\end{table}

The numerical values extracted from the best fits, as well as the value of
the peak frequency of $I_{tot}$, 
$\omega_B$, are reported in Table II. The fits are displayed in Fig. 4 
and they show that the closer $\omega_B \tau$ is to unity, the worse
is the agreement with the density-only model. Note that the 
value of $2\Lambda' a/3 \rho_m b$ has been overestimated to 
emphasize the effect. The discrepancy between the fit function 
and the computed spectra has always the characteristic features 
discussed before: the fit function has a lower intensity than the
computed one above the Brillouin peak and a higher one below. 

The trends reported in Table II are worth commenting: the ratio $\tau_L/\tau$
increases with longer relaxation times $\tau$, i.e. lower temperatures: 
the apparent relaxation time $\tau_L$ is longer than the 
original one. In Brillouin experiments on supercooled liquids, this effect 
will take place for relaxation times
of the order of some nanoseconds, while it cannot
be identified for much longer relaxation times. The effect will
then be apparent in the region where the curve $\log \tau$ versus $1/T$ has 
its maximum curvature for fragile glass forming liquids. This curvature is 
the origin of the nonzero value of the Vogel-Fulcher
temperature: an apparent increase of the relaxation times
in this time window will decrease this curvature and result in an artificial
decrease of the Vogel-Fulcher temperature.

Table II also exhibits an effect that has not been reported in
experimental studies of Brillouin spectra
 of supercooled liquids: a decrease,
instead of an increase,
of $\Delta_L$ upon cooling 
\cite{Dreyfus1992,Du1994,Aouadi2000,Shen2000,Monaco2001}. 
This contrasts with the 
familiar trend of $\omega_L$ which increases with $\tau$ as is shown 
in Table II. The unconventional dependence on temperature of 
$\Delta_L$ may be an artefact of the oversimplified model for the 
memory kernels.

\section{Summary and final remarks}

The present paper aimed at giving  complete and transparent 
expressions for polarised
Brillouin light scattering experiments  within the
framework of the phenomenological equations recalled in the Introduction.
These equations
have been derived through the use of heuristic arguments in [I] and a 
demonstration of their validity has been briefly sketched in \cite{Latz2001}. 
Their detailed proof will be given in the second paper of this 
series (Part II), where the same set of relevant variables, i.e
mass density $\rho$, orientation $Q_{ij}$ and their respective 
currents are considered. 

The expressions for the intensity in the VV and HH scattering geometry
are given by Eqs. (\ref{IVV}) and (\ref{IHH}). In both cases, the spectrum
naturally splits into a sum of two terms: one describes the pure orientational
dynamics of the molecules, decoupled from density fluctuations; the
second term involves the propagation of longitudinal phonons. 
The corresponding longitudinal viscosity  comprises contributions from 
 the relaxation 
of the translational and orientational motions. The coupling of these
two types of motion is characterised by the translation-rotation coupling
constant, $\Lambda'$, and by 
the frequency-dependent rotation-translation
function, $\mu(\omega)$, while  the orientational dynamics 
is characterised by 
$D(\omega)$ which depends very weakly on $\omega$ in the region of 
interest for Brillouin scattering studies. 
The role of $\Lambda'$ and $r(\omega)$ in the 
 spectra is twofold: first, their play a role in  the phonon propagator 
and, second,  they enter in the detection mechanism via the 
factor $[ a + 
2\Lambda' b r(\omega) /3 \rho_m ]$. 

The role of the molecular polarisability anisotropy in detecting both the 
uncoupled orientational dynamics and the transverse, diffusive or propagative
modes with wave vector $\vek{q}$ already appeared in [I]. For the
sake of completeness, we reproduce here the
results obtained in that paper under a from that allows for an easy 
comparison with \gl{IVV}:
\begin{eqnarray}\label{IVH}
& & I_{VH}(q,\omega) =  \frac{\langle | Q_{\perp\perp'}^0 |^2 \rangle 
}{\omega} Im \Bigg\{
b^2 \left[ 1-\frac{\omega_0^2}{D(\omega)}\right] \nonumber \\
& & + \frac{\rho_m}{\Lambda'} 
(\omega_o q)^2 \cos^2 \frac{\theta}{2} P_T(q,\omega) \left[
\frac{\Lambda'}{\rho_m} b r(\omega) \right]^2 \Bigg\} 
\, ,
\end{eqnarray} 
where the 
$q$-dependent part is now mediated by the transverse phonon propagator:
\begin{eqnarray}
P_T(q,\omega) = \left[ \omega^2-  q^2 \rho_m^{-1} \omega \eta_T(\omega)
\right]^{-1} \, ,
\end{eqnarray}
characterised by the transverse viscosity:
\begin{equation}
\eta_T(\omega) = 
 \eta_s(\omega)
- \frac{\Lambda'}{\omega} D(\omega) r(\omega)^2\, .
\end{equation}
The absence, in the second 
square bracket of \gl{IVH} of the factor 2/3 is in 
agreement
with a result of \cite{Franosch2001}, as will be discussed in Part II.

We studied, in Section III, the change in the spectral line-shape brought by
taking into account the additional light scattering channels. We concentrated
 on the second term of the r.h.s. of \gl{IVV}, mostly discussing the type of 
distortion produced by these new terms in the  spectral shape in the vicinity
of the Brillouin peak. Let us add some remarks.

Firstly, in Section III, the numerical calculation of the total intensity
was performed using a positive ratio $b/a$ and we assumed
through \gl{single} that $\mu(t)$ could be characterised
by some amplitude, $\Delta_\mu^2$, and a smooth, positive, decreasing
function of time that approaches zero for long times. 
Contrary to the coefficient $a$ in \gl{eps}, 
the numbers $b$ and $\Delta_\mu^2$ are not always positive, but we want 
to point out that their product $b\Delta_\mu^2$ is commonly a positive
quantity, for prolate as well as for oblate axial molecules. 
Indeed, on the one hand, the polarisability anisotropy, $b$, is usually
positive for prolate molecules giving rise to glass-forming liquids
 and negative for the oblate ones. On the other hand, the sign of 
$\mu(t)$ is also shape dependent as can be inferred from the following
argument. Subjected to a steady shear flow, $\tau_{ij}$,  a molecular
liquid builds up a nonzero stationary mean orientation $Q_{ij}$
which is easily deduced from \gl{Qosci}: 
\begin{eqnarray}
Q_{ij} = \frac{1}{\omega_0^2} \left[ \int_0^\infty \mu(t) dt \right]
\tau_{ij} \, .
\end{eqnarray}
If, for instance, this flow is in the $\hat{\vek{x}}$ 
direction with a positive
gradient
  in the $\hat{\vek{z}}$ direction,  i.e. $\tau_{xz}  > 0$, 
one easily convinces oneself
that a long axis of the molecule 
will be, on average, parallel to $\hat{\vek{x}} 
+ \hat{\vek{z}}$, and a short axis parallel to 
$ \hat{\vek{x}} - \hat{\vek{z}}$. Hence, cigar-shaped molecules will 
exhibit $Q_{xz}  >0$, whereas disk-shaped ones lead to 
$Q_{xz}  < 0$. In other words, the shear flow exerts a torque
on the molecules so that the sign of:
\begin{equation}
\int_0^\infty \mu(t) dt = Im \mu(\omega=0) \, ,
\end{equation}
which is also the sign of $\Delta_\mu^2$ for a smoothly decreasing function
$Im \mu(\omega)$,
is the same as the sign of $b$.

Secondly, the total VV spectrum is the sum of two terms in \gl{IVV}: the 
first is  proportional to the backscattering VH spectrum and gives always
a positive contribution to the intensity for all frequencies. Conversely, 
at very low frequencies, the second, $q$-dependent, term may give rise to  a 
negative contribution. This is in contrast to the density-only model and may
serve as a simple test to determine whether or not translation-orientation
coupling is significant. Since the measured spectrum should be
 positive whatever 
the frequency,  
 one would like to know if the proposed phenomenological equations
ensure this property and, if necessary, what additional requirements have
to be imposed on the memory kernels 
to guarantee this positiveness.
 The phenomenological equations allow to derive the
appropriate conditions. However, these conditions
 will appear more naturally and in a 
transparent way in the 
 microscopic approach. Therefore their derivation will be postponed to
Part II and we merely state here the result: the  measurable spectra
 in the three geometries VV, HH, VH, Eqs. (\ref{IVV}) (\ref{IHH}) and (\ref{IVH}), 
are positive for all frequencies provided that: 

a) the imaginary part of $\Gamma'(\omega), \eta_b(\omega)$ and
$\eta_s(\omega)$ are positive for all frequencies;

b) the imaginary part of the translation-rotation coupling fulfills:
\begin{equation}\label{Onsager}
[Im \Gamma'(\omega)] [Im \eta_s(\omega)] - \Lambda' [Im \mu(\omega)]^2 > 0
\, .
\end{equation}

This conditions appear as a generalisation of the Onsager relations 
for the dynamics of a coupled system and are, as expected, independent of the
$a/b$ ratio. This corroborates the argument that the positiveness of the 
spectra should not depend on the relative strength of the 
two scattering mechanisms.  In view of the analytical form of the 
memory kernels as given in \gl{single}, and with the numerical values used
in Section III
for $\Delta_{\eta_b}, \Delta_{\eta_s}, \Delta_{\mu}$ and $\Lambda'$, 
\gl{Onsager} is fulfilled for all frequencies if $\Delta_{\Gamma'}^2 > 1/2$. 
Nevertheless, in the frequency range considered, the role of $\Delta_{\Gamma'}$
is important only in the $q$-independent part of \gl{IVV} (rotational part of 
the spectrum):
\begin{eqnarray}\label{ImD}
\frac{1}{\omega} Im \left[ 1- \frac{\omega_0^2}{D(\omega)} \right]
= \frac{1}{\omega} Im \left[ 
\frac{\omega \Gamma'(\omega) - \omega^2}{D(\omega)} \right] \, .
\end{eqnarray} 
 We have checked that the neglect of $\Delta_{\Gamma'}$ 
in $D(\omega)$, both in the denominator of the r.h.s. of \gl{ImD} 
and in $r(\omega)$, has virtually no influence:
it does not change the intensity and the shape, either  of the
pure
 rotational spectrum, or of the $q$-dependent spectrum 
 discussed in Section III. 
Conversely, this neglect allows for the analytic discussion 
of $I_{tot}$ performed in that
section. This justifies, a posteriori, the simplification made there. 

Finally, we already pointed out in the Introduction that effects related 
to energy conservation, in particular to thermal diffusion, are neglected
in the present paper. 
Some consequences of the heat diffusion process are well known since the
pioneer work of Landau and Placzek \cite{Landau1934}. 
It gives rise to contributions
in the propagator of longitudinal phonons for very low frequencies. 
Other aspects of the role of temperature fluctuations were taken into
account in \cite{Franosch2001}. 
There, it was shown that, if one deals explicitly 
with the coupling of temperature to dielectric fluctuations, one arrives 
at a rather complex form of the coupling of detection channels and
propagators of these excitations. In order to describe this 
aspect within a phenomenological approach, the constitutive equations
of the present paper have to be generalised and supplemented by an 
equation of motion for  the energy conservation, while the detection 
mechanism, \gl{eps}, will remain unchanged.

\begin{acknowledgments}
This paper has benefited from many discussions of one of us (R.M.P)
with  H.Z. Cummins on the light scattering mechanisms. The latter was
also instrumental in suggesting the collaboration that lead to the present
paper. 
\end{acknowledgments}

\end{document}